\documentclass[aps,prb,preprint,showpacs,groupedaddress]{revtex4}

\usepackage{graphicx}
\usepackage{epsfig}
\preprint{lao-sto-2017}
\usepackage{color} 

\begin{document}
\DeclareGraphicsExtensions{.eps,.jpg,.png}
\input epsf
\title{Infrared study of the quasi-two-dimensional electron system at the interface between SrTiO$_{3}$ and crystalline or amorphous LaAlO$_3$} 
\author {A. Nucara$^{1}$, M. Corasaniti$^{2}$, A. Kalaboukhov$^{3}$, M. Ortolani$^{2}$,  E. Falsetti$^{2}$,  A. Sambri$^{4}$, F. Miletto Granozio$^{4}$,  F. Capitani$^{5}$, J.-B. Brubach$^{5}$, P. Roy$^{5}$, U. Schade$^{6}$, and P. Calvani$^{1}$}
\affiliation{$^{1}$CNR-SPIN and Dipartimento di Fisica,  Universit\`{a} di Roma ''La Sapienza'', P.le A. Moro 2, I-00185 Roma, Italy}\
\affiliation{$^{2}$Dipartimento di Fisica,  Universit\`{a} di Roma La Sapienza, P.le A. Moro 2, I-00185 Roma, Italy}\
\affiliation{$^{3}$Department of Microtechnology \& Nanoscience, Chalmers University, S-41296 Gothenburg, Sweden}\
\affiliation{$^{4}$CNR-SPIN UOS Napoli, Complesso Universitario di Monte Sant'Angelo, Via Cinthia I-80126, Napoli, Italy}
\affiliation{$^{5}$Synchrotron SOLEIL, L'Orme des Merisiers Saint-Aubin, BP 48, F-91192 Gif-sur-Yvette Cedex, France}\
\affiliation{$^{6}$Helmholtz Zentrum Berlin Mat \& Energie GmbH, Methods Mat. Dev., D-12489 Berlin, Germany}\

\begin{abstract}
We have used grazing-angle infrared spectroscopy to detect the Berreman effect (BE) in the quasi-two-dimensional electron system (q-2DES) which forms spontaneously at the interface between SrTiO$_{3}$ (STO) and a thin film of LaAlO$_3$ (LAO). From the BE, which allows one to study longitudinal optical excitations in ultrathin films like the q-2DES,  we have extracted at different temperatures its thickness,  the charge density and mobility of the carriers under crystalline  LAO (sample A), and the charge density under amorphous LAO (sample B). This quantity turns out to be higher than in sample A, but a comparison with Hall measurements shows that under amorphous LAO the charges are partly localized at low $T$ with a low activation energy (about 190 K in $k_B$ units), and are thermally activated according to a model for large polarons. The thickness of the q-2DES extracted from our spectra turns out to be 4 $\pm 1$ nm for crystalline LAO, 7 $\pm 2$ nm for amorphous LAO. 
\end{abstract}

\maketitle

\section{Introduction}
The interest for two-dimensional electron systems (2DES) dates back, at least, to the observation of Wigner electron crystals on the surface of superfluid helium \cite{Sommer}. However, the studies on the 2DES have reached their widest diffusion in solid state physics, through the fabrication of  semiconductor-based heterostructures for the implementation of electronic and photonic devices. In the last decade, moreover, it has been discovered that a 2DES may  form spontaneously in systems of extraordinary interest like  graphene \cite{Geim}, topological insulators \cite{Hasan,Moore}, and the interfaces between insulating oxides. The most studied 2DES in oxides is the free-electron layer which forms within the upper TiO$_2$ layer of a SrTiO$_3$ (STO) substrate, when a LaAlO$_3$ (LAO) film thicker than 4 unit cells \cite{Ohtomo,Schlom} is deposited on it. The  thickness of this electron layer is not larger than 10 nm  \cite{Basletic,Dubroka} and for this reason is often called a quasi-2DES (q-2DES). Among the intriguing properties of the LAO/STO q-2DES, both ferromagnetism at low temperatures \cite{Brinkman}, and superconductivity below 200 mK \cite{Reyren} have been reported, two phenomena which can also coexist \cite{Li,Bert}. Similar two-dimensional electron systems have been also observed in NdGaO$_3$/SrTiO$_{3}$  \cite{DiGennaro}, LaAlO$_3$/EuTiO$_{3}$/SrTiO$_{3}$ \cite{DeLuca}.

The q-2DES formation is basically explained in terms of a top-down charge transfer aimed at preventing a "polar catastrophe"  within LAO \cite{Ohtomo,Caviglia}.  Nevertheless, several authors have stressed the role  of the oxygen vacancies \cite{Wilmott,Alexei} which may form during the film growth, survive to the standard annealing procedures, and dope the interface by electrons. The latter mechanism may better explain why a 2DES similar to that of crystalline LAO is observed also under amorphous LAO \cite{Chen}.  
While the dc conducting properties of the 2DES in LAO/STO have been determined since the beginning \cite{Pentcheva}, to our knowledge only a couple of experiments have been devoted to studying its low-energy electrodynamics \cite{Dubroka,Dubroka17}. In fact, determining the frequency-dependent response of ultrathin metallic films is a challenging task for conventional infrared (IR) spectroscopy. Both the above cited experiments indeed obtained reliable results by exploiting an important effect discovered by D. W. Berreman  in 1963 \cite{Berreman}. 

In the Berreman effect (BE), the $p$-component of the electric field (in the plane of incidence) undergoes enhanced absorption through a film of thickness  $d$, if $d$ is much smaller than both the radiation wavelength $\lambda$ and the field penetration depth $d_p$. The BE thus allows one to investigate the low-energy electrodynamics of very thin films.  However, it occurs only at  frequencies $\omega$ close to that  where the real part of the film dielectric function $\epsilon_1(\omega)$ vanishes. Consider for sake of simplicity a film (index $f$) self supporting in a vacuum (index 0). If the angle of incidence $\theta$ is large, the $p$-polarized  component of the incident electric field is strong and enters the film normal to its surface ($E_n^0(\omega)$). Therein, it becomes $E_n^f(\omega)= D_n^f(\omega)/\epsilon_1(\omega)$, where $\epsilon_1(\omega)$ is the real part of the dielectric function and $D_n^f(\omega)$ is the normal component of the electric displacement field. As $D_n^f = D_n^{0} = E_n^0$ for the continuity equations, at $\omega$'s such that $\epsilon_1(\omega) \simeq 0$ one has both a pole in the energy loss function and a strong field enhancement ($E_n^f(\omega) >> E_n^0(\omega)$). Thus, in a film thinner than both $\lambda$ and $d_p$, $E_n(\omega)$ creates a dynamical dipole moment which resonates with the longitudinal excitations of the sample \cite{Harbecke}. The same mechanism works in thin films deposited on a substrate, provided that suitable formulas (see Section III) are used to take into account the optical response of the various interfaces. The excitations detectable in a  BE experiment include the vibrations of a thin layer of adsorbed molecules, the longitudinal optical phonons of insulating thin films (at frequencies $\omega_{L}$), and the screened plasma frequency $\omega_p/\sqrt{\epsilon_{\infty}}$ of metallic  films (where $\epsilon_{\infty}$ is the real part of the dielectric function for $\omega >> \omega_p$). 

Object of the present investigation is indeed the response to grazing-angle incident radiation of the few-nm thick q-2DES at the interface between LaAlO$_3$ (either crystalline or amorphous) and SrTiO$_{3}$.  Unlike in previous similar experiments \cite{Dubroka,Dubroka17} we did not use a real ellipsometric apparatus, but we measured at grazing angle, by using a single polarizer and synchrotron radiation, the ratio $R_p/R_s$ where $R_p$ and $R_s$ are the LAO/STO reflectances in $p$ and $s$ polarization, respectively. Afterwards, in order to compare our results with those in the literature, we have determined the ellipsometric angle 

\begin{equation}
\Psi(\omega) = \arctan (R_p/R_s)^{1/2} = \arctan[|\tilde r^p(\omega)|/|\tilde r^s(\omega)|] 
\label{Psi}
\end{equation}

\noindent
where $\tilde r^p$ and $\tilde r^s$  are the complex reflection coefficients in the two polarizations\cite{Born}.  In order to measure $R_p/R_s$ one has just to remotely turn a polarizer in front of the sample, and then to repeat the same operation after shifting the mirror to the sample position,  without acting on  the optics or on the sample. Moreover, in order to check that the observed BE comes from the q-2DES, we have measured $\Psi(\omega)$ in a third LAO/STO sample - where the q-2DES was erased by ion etching \cite{Nucara} - that was used as reference. This will allow us to extract the predicted shape of the Berreman resonance \cite{Dubroka}, which consists of a dip-peak feature in the difference between the angles $\Psi$ measured in the sample with the q-2DES and in the no-q-2DES reference. By fitting to data the optical formulas reported in Section III we have then determined  the electrodynamic parameters of the q-2DES for both crystalline and amorphous LAO. In the former case we substantially confirm the results of  Ref. \onlinecite{Dubroka}, by a somewhat simpler procedure, while the present measurements on amorphous LAO are the first ones of this kind to our knowledge. They show a q-2DES charge density remarkably higher than under crystalline LAO, but a comparison with Hall measurements shows that the charges are partly localized at low $T$ with a low activation energy (about 50 K in $k_B$ units), suggesting that they may form large polarons.

\section{Experiment and results}
Two  crystalline LAO films (samples A and C), and an amorphous-LAO film (sample B) were deposited on TiO$_2$-terminated SrTiO$_3$ substrates by Pulsed Laser Deposition (PLD). In samples A and C, LAO was 10 unit cells (4 nm) thick, so that it could develop the 2DES at the interface with STO while its absorption was negligible \cite{Dubroka,Nucara}. They were grown \cite{Aurino} at a substrate temperature of 800 $^\circ$C and under an oxygen pressure of 10$^{-4}$ mbar, then further annealed for one hour at 600 $^\circ$C and 500 mbar.  The sheet resistance of the conducting 2DES at the interface has been reported in Ref. \onlinecite{Nucara}. Sample B was grown at room temperature by pulsed laser deposition using a KrF excimer laser, at an oxygen pressure of 5$\times$10$^{-4}$ mbar. The LAO thickness was 5nm, based on calibrations made on the RHEED oscillations of a twin crystalline sample grown at room temperature. The sheet resistance of sample B is reported in Fig. 1 of the Supplementary Information (SI) \cite{SI}. Afterwards, the 2DES in sample C  was erased by ion milling dry etching as described in Ref. \onlinecite{Aurino}, where the procedure was shown to damage LAO/STO within a depth of just 1 nm. The low-energy ion irradiation does not etch the LAO film and the morphological and chemical properties of sample C are very similar to sample A, but without conducting layer. It was measured in the same way as samples A and B, to be used as reference.

\begin{figure} 
\begin{center}
\resizebox{1.2\columnwidth}{!}
{\includegraphics{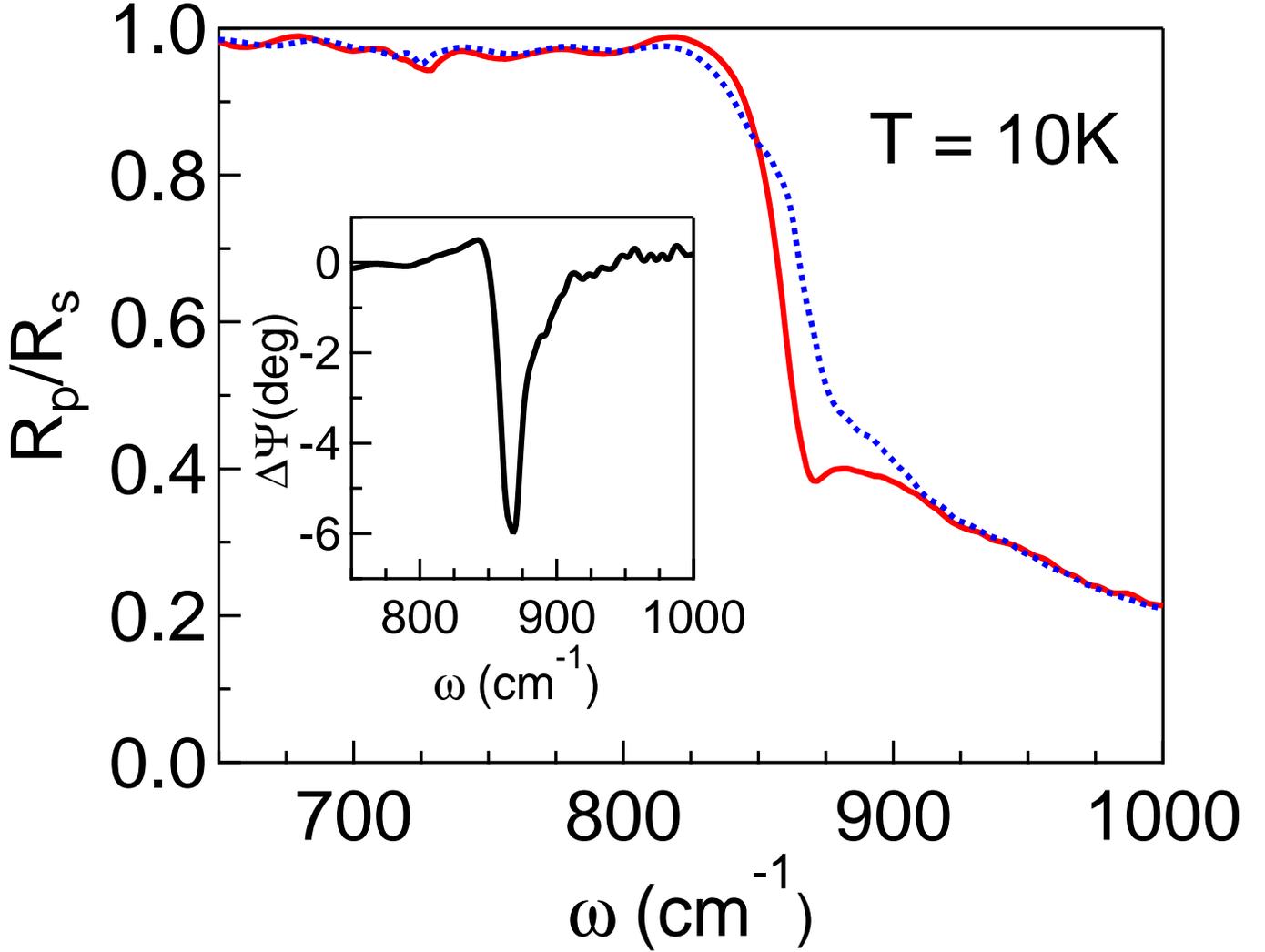}}
\caption{Ratio $R_p/R_s$ measured at the beamline IRIS of BESSY-II in the amorphous-LAO/STO sample B with the q-2DES (solid red line) and in the LAO/STO sample C without q-2DES (dotted blue line) at 10 K. The inset shows the difference $\Delta \Psi = \Psi_B - \Psi_C$  between the ellipsometric angles calculated by Eq. \ref{Psi} for B and C. The profound dip at 868 cm$^{-1}$ is  caused by the Berreman effect in the q-2DES.} 
\label{Psi-amorph} 
\end{center}      
\end{figure}

\begin{figure} 
\begin{center}
\resizebox{1.0\columnwidth}{!}
{\includegraphics{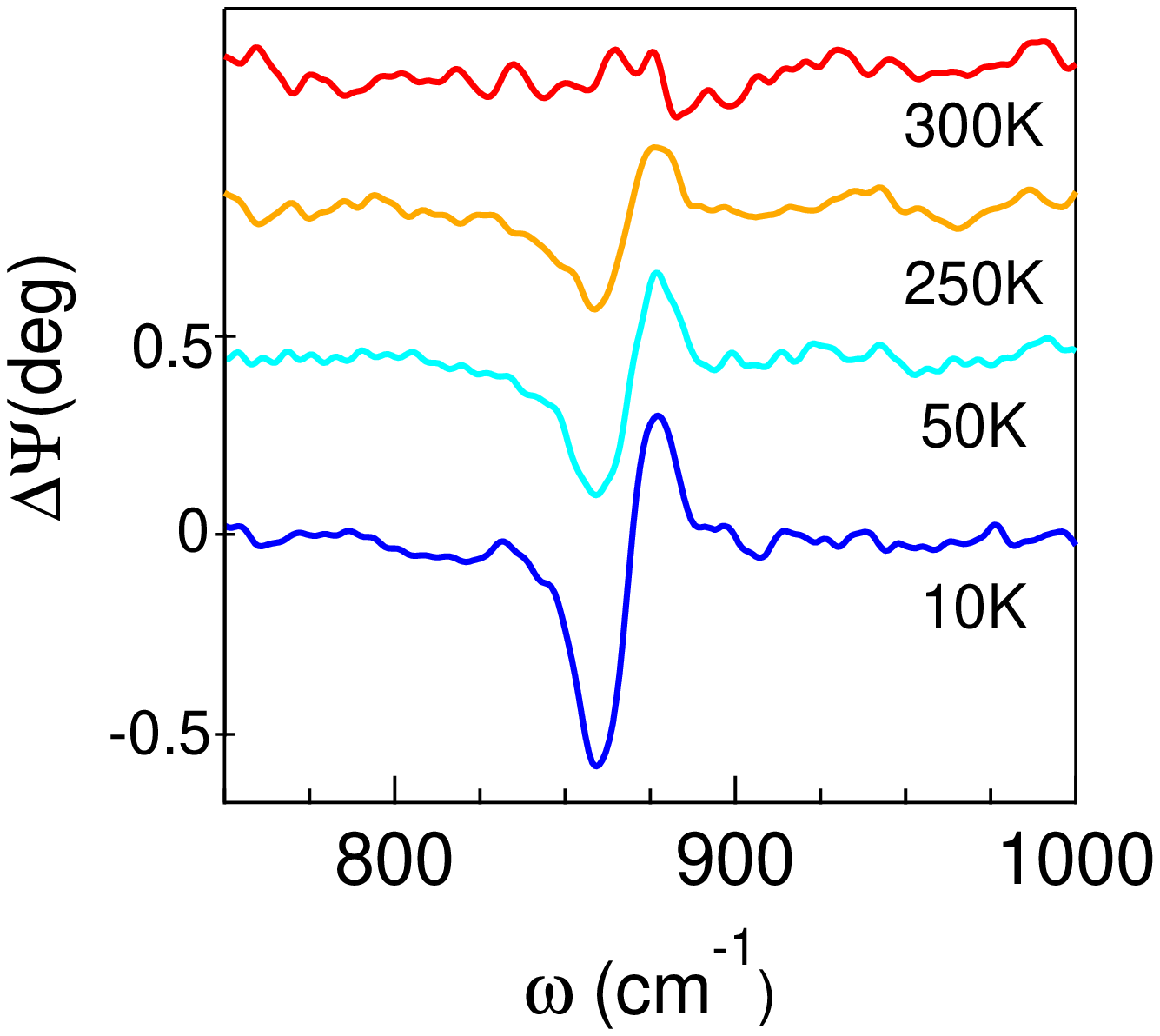}}
\caption{Berreman effect caused by the q-2DES at the interface between crystalline LAO and STO, as obtained by subtracting, from the $\Psi(\omega)$ of sample A, that  measured in  LAO/STO  without q-2DES (sample C). The vertical scale reported for T = 10 K holds also for the other temperatures where the zeros have been shifted for clarity. Data were taken at the beamline AILES of SOLEIL.} 
\label{crystal} 
\end{center}      
\end{figure}

The reflectivity $R_j(\omega)$ of  samples A, B, and C  was measured in both polarizations $j=p,s$ under an angle of incidence $\theta = 72^\circ$, by Michelson interferometers and Mercury-Cadmium Telluride (MCT) detectors cooled by liquid nitrogen. In order to improve the signal-to-noise ratio at grazing incidence, we have exploited the higher brilliance of infrared synchrotron radiation with respect to conventional black bodies \cite{Roy}.  Preliminary measurements at the beamline  IRIS of BESSY-II were repeated and completed at the beamline AILES of SOLEIL by a different apparatus. In both experiments the samples were thermoregulated within $\pm$ 2 K. Golden mirrors placed above the sample and aligned parallel to it by a laser beam were used as references. An external mechanism allowed one to illuminate either the sample or the mirror. A single KRS-5 polarizer having a contrast better than 99.9 \%, that could be remotely rotated, was put on the radiation path. Optical filters were used in order to confine the  spectrum of the incident radiation within  the STO gap (3.2 eV), thus preventing any possible photodoping of the interface. At each temperature, two 300-scan series of interferograms were taken on the sample, both in $p$ and $s$ polarization, with a spectral resolution of 2 cm$^{-1}$. The same sequence was then repeated at each temperature for the Au mirror. We thus obtained the reflectances $R_p$ and $R_s$ shown in Fig. 2 of the SI \cite{SI} and, by Eq. \ref{Psi}, the ellipsometric angle $\Psi(\omega)$ for samples A, B, and C.

The ratios $R_p/R_s$ measured at BESSY-II in sample B, which has the q-2DES, and in sample C, where the q-2DES had been erased, are shown in Fig. \ref{Psi-amorph} at 10 K in the region of the hardest longitudinal phonon of STO (L3).  A strong anomaly appears above its frequency $\omega_{L3}$ =790 cm$^{-1}$ in the former spectrum only. This comparison shows that the anomaly is fully associated with the q-2DES and that sample C, quite similar to the other ones except for the first nanometer-thick layer from the surface, can be used as reference in a sort of differential spectroscopy. The inset shows indeed the difference $\Delta \Psi = \Psi_B - \Psi_C$  between the ellipsometric angles calculated by Eq. \ref{Psi} for B and C. A similar procedure was used in Ref. \onlinecite{Dubroka}, where however the reference was a bare STO substrate. The resulting dip at 868 cm$^{-1}$, about 70 cm$^{-1}$ from $\omega_{L3}$, is characteristic of the Berreman effect and can be entirely ascribed to the q-2DES. Such dip is much stronger than that observed previously \cite{Dubroka,Dubroka17} and also in the present experiment (see below) under crystalline LAO. 

\begin{figure} 
\begin{center}
\resizebox{0.9\columnwidth}{!}
{\includegraphics{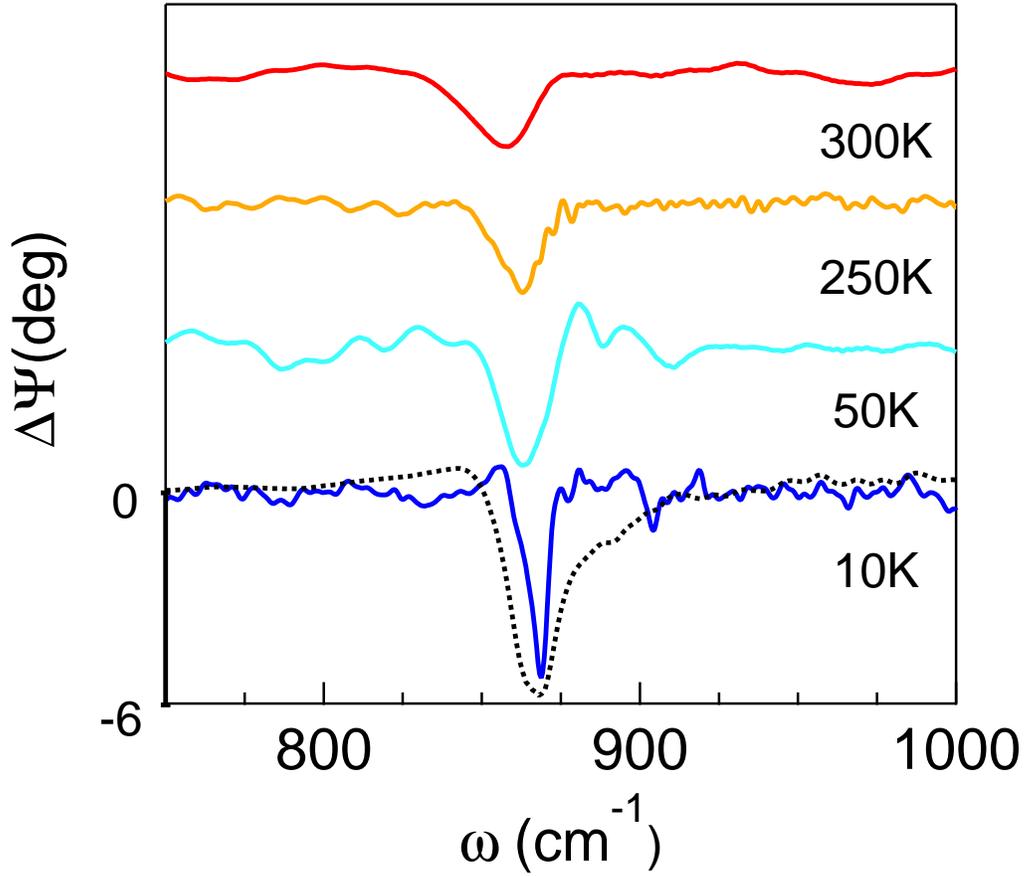}}
\caption{Berreman effect caused by the q-2DES at the interface between amorphous LAO and STO, as obtained by subtracting, from the $\Psi(\omega)$ of sample B, that  measured in  LAO/STO  without q-2DES (sample C). The vertical scale reported for T = 10 K holds also for the other temperatures where the zeros have been shifted for clarity. Data were taken at the beamline AILES of SOLEIL. The corresponding data obtained at BESSY-II at 10 K on the same sample but with a different optical setup are reported for comparison by a dashed line on the same scale (see text).} 
\label{amorph} 
\end{center}      
\end{figure}

Figure \ref{crystal} shows the results of the same experiment when applied to crystalline LAO on STO (sample A) at different temperatures. In $\Delta \Psi = \Psi_A - \Psi_C$, a full  Berreman resonance is observed,  consisting of a dip at 859 cm$^{-1}$ followed by a peak at $\omega_{Ber} =$ 875 cm$^{-1}$, as observed in Ref. \onlinecite{Dubroka} for crystalline LAO/STO. The peak  is shifted by about 85 cm$^{-1}$ from the $\omega_{L3}$ of STO. The fact that $\omega_{Ber}$  is independent of temperature is consistent with both the prediction that the peak frequency is determined essentially by the carrier density  in the q-2DES\cite{Dubroka,Dubroka17} , and that this latter has a really metallic behavior in the infrared. 

Figure \ref{amorph} displays instead $\Delta \Psi = \Psi_B - \Psi_C$ for the q-2DES under amorphous LAO at different temperatures, as obtained from data collected at SOLEIL. At 10 K, those taken at BESSY-II are reproposed for comparison by the dotted line.  All spectra in Fig. \ref{amorph} show a Berreman dip much more pronounced than in Fig. \ref{crystal}, while the Berreman peak at  $\omega_{Ber}$ is not observed. This point will be discussed in the next Section. The $\Delta \Psi$ obtained  at 10 K with the two different apparata provide a Berreman dip at the same frequency and with the same amplitude within errors. However,  the BESSY-II source  provides on one hand a better signal-to-noise ratio, on the other hand, a larger dip linewidth. As it results from the calculations of Ref. \onlinecite{Park}, the dip width has no particular meaning, being not related to the carrier relaxation time which instead affects the Berreman peak. Here, we attribute the width mismatch between the two apparata to the helium-flow cryostat windows, which at BESSY-II may slightly mix the $p$ and $s$ components after the polarizer. At SOLEIL there are no such windows, as the sample is cooled by a cryogenerator and the whole optical chamber, including the polarizer, is in a high vacuum.

\section{Data analysis and discussion}

\begin{figure} 
\begin{center}
\resizebox{0.9\columnwidth}{!}
{\includegraphics{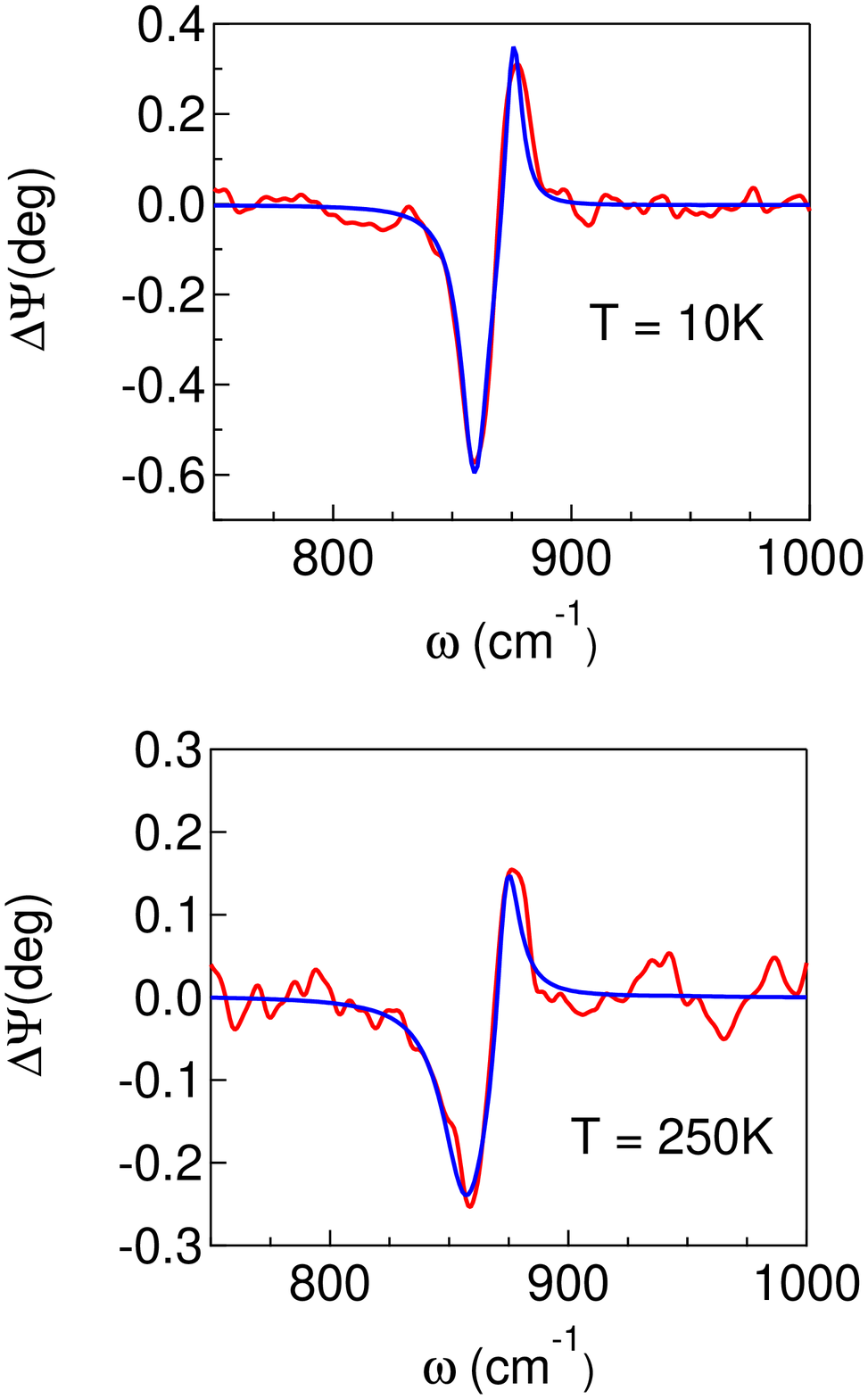}}
\caption{Fits  at two temperatures by use of Eqs. \ref{Psi} to \ref{LST+D}  (blue lines) to the experimental $\Delta \Psi = \Psi_A - \Psi_C$  of Fig. \ref{crystal} (red lines) for crystalline LAO on STO.} 
\label{fitA} 
\end{center}      
\end{figure}

The analysis of the spectra proceeded as follows. We assume that the 4 nm thick LAO is transparent at all frequencies, and that the STO substrate with thickness $d$ = 0.5 mm is a semi-infinite medium. The latter hypothesis has been checked by fitting the reflectivity of the no-2DES sample C with $d$ as a free parameter.  We then modeled the samples with the 2DES by  an optical double layer \cite{Calvani1} formed by a conducting STO film of thickness $d$ and refraction index $\tilde n_1$, which contains the free charges, and by the STO substrate of index $\tilde n_2$. Two interfaces of the q-2DES are both transmitting and reflecting, the one with a vacuum (indexed as 01)  and the one with the insulating STO  (indexed as 12). One then has for the reflection coefficients of the entire multilayer\cite{Dressel} 

\begin{equation}
\tilde r^j_{012} = \frac{\tilde r^j_{01}+\tilde r^j_{12}exp(2i \delta)}{1+\tilde r^j_{01}\tilde r^j_{12}exp(2i \delta)}
\label{coeff}
\end{equation}

\noindent
where $j = p, s$ identifies the polarization,  $\delta = 2 \pi \tilde n_1/\lambda_0$, and $\lambda_0$ is the radiation wavelength in a vacuum. For sample C, $r^j_{12}$ = 0 and the reflectivity reduces to that of bulk STO.

In Eq. \ref{coeff}, the reflection coefficients at the interface with a vacuum ($\tilde n_0$ =1)  for an angle of incidence $\theta$ (here 72$^\circ$), are given by the Fresnel formulas \cite{Dressel}

\begin{eqnarray}
\tilde r^p_{01} = \frac{\tilde n^2 cos \theta - \sqrt{\tilde n^2 - sin^2 \theta}}{\tilde n^2 cos \theta + \sqrt{\tilde n^2 -sin^2 \theta}} \\ 
\tilde r^s_{01} = \frac{cos \theta - \sqrt{\tilde n^2 - sin^2 \theta}}{cos \theta + \sqrt{\tilde n^2 -sin^2 \theta}} \nonumber
\label{Fresnel1}
\end{eqnarray}

\noindent
where $\tilde n = \tilde n_1$  for the conducting samples A,B and $\tilde n = \tilde n_2$ for the insulating STO sample C.  In turn, the reflection coefficients at the  interface between conducting and insulating STO are obtained by 

\begin{eqnarray}
\tilde r^p_{12} = \frac{(\tilde n_2/\tilde n_1)^2 cos \theta - \sqrt{(\tilde n_2/\tilde n_1)^2 - sin^2 \theta}}{(\tilde n_2/\tilde n_1)^2 cos \theta + \sqrt{(\tilde n_2/\tilde n_1)^2 -sin^2 \theta}} \\ 
\tilde r^s_{12} = \frac{cos \theta - \sqrt{(\tilde n_2/\tilde n_1)^2 - sin^2 \theta}}{cos \theta + \sqrt{(\tilde n_2/\tilde n_1)^2 -sin^2 \theta}} \nonumber
\label{Fresnel}
\end{eqnarray}

The refraction index for the no-2DES sample, $\tilde n_2 = \sqrt{\tilde \epsilon_2}$, is obtained by the Lyddane-Sachs-Teller expression of the complex dielectric function for STO \cite{Gervais} 

\begin{equation}
\widetilde{\epsilon_2}(\omega)=\epsilon_{\infty}\prod_{j}\frac{(\Omega_{Lj}^{2})-\omega^2+\textit{i}(\Gamma_{Lj})\omega}{(\Omega_{Tj}^{2})-\omega^2+\textit{i}(\Gamma_{Tj})\omega}
\label{LST}
\end{equation}

\noindent
where $\Omega_{j}$ and  $\Gamma_{j}$ are the central frequency and  width, respectively, of either the \textit{j-th} longitudinal (L) or transverse (T) optical phonon. 
In turn, the refraction index in the presence of the q-2DES $\tilde n_1 = \sqrt{\tilde \epsilon_1}$  is calculated by adding to Eq. \ref{LST} a free-electron Drude term, so that  

\begin{equation}
\widetilde{\epsilon_1}(\omega)= \widetilde{\epsilon_2}(\omega) + \widetilde{\epsilon_D}(\omega) = \widetilde{\epsilon_2}(\omega)  - \frac{\omega_p^2}{\omega^{2}+i\omega\Gamma_D}
\label{LST+D}
\end{equation}

\begin{figure} 
\begin{center}
\resizebox{0.9\columnwidth}{!}
{\includegraphics{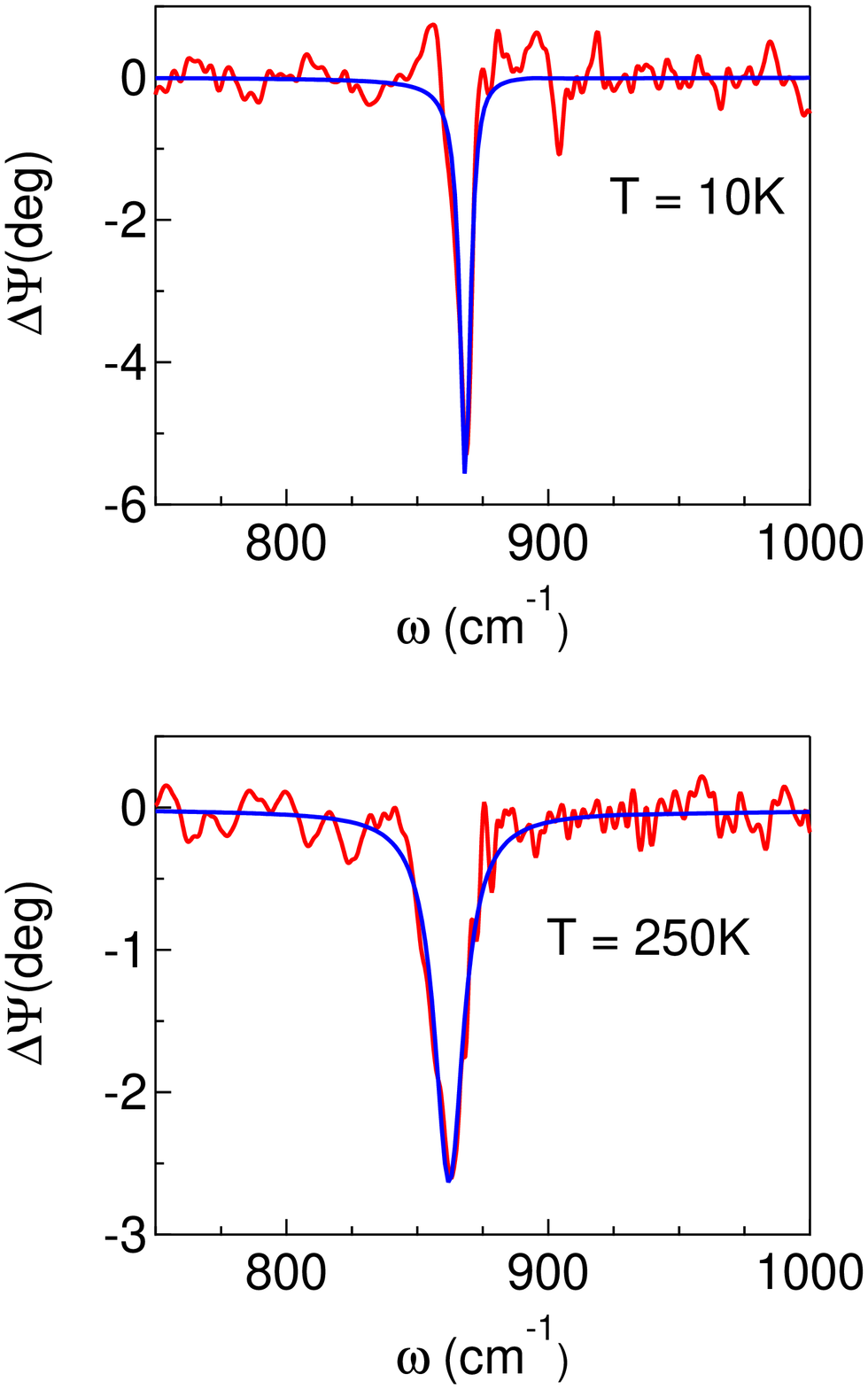}}
\caption{Experimental $\Delta \Psi = \Psi_B - \Psi_C$ (red lines), from data taken at the beamline AILES of SOLEIL, showing the Berreman effect in amorphous LAO on STO at two temperatures. The blue lines are the fits  to data obtained from Eqs. \ref{Psi} to \ref{LST+D}. } 
\label{fitB} 
\end{center}      
\end{figure}

The curves obtained  by fitting to data Eqs. \ref{Psi} to \ref{LST+D} are shown at different temperatures in Figs. \ref{fitA} and \ref{fitB} for crystalline and amorphous LAO, respectively. Despite the numerous equations involved and the weakness of the Berreman effect in the q-2DES, the fits are quite satisfactory. The resulting parameters of the Drude-Lorentz model for the conducting film are reported in Table I. The other phonon frequencies and widths were  fixed to the values reported in Table II of Ref. \onlinecite{Nucara}. The q-2DES thickness is independent of temperature within errors and turns out to be 4 $\pm$ 1 nm under crystalline LAO, consistently with the direct determination of Ref. \onlinecite{Basletic} ($d < 7$ nm), but increases to 7 $\pm$ 2 nm under amorphous LAO.

\begin{table} 
\caption{Fitting parameters for the q-2DES at the interface between crystalline (A) or amorphous (B) LAO and STO. The plasma frequency $\omega_p$, the relaxation rate $\Gamma_D$, the highest LO-phonon frequency and width for STO, are all expressed in cm$^{-1}$. For the  parameters of the other STO phonons we used the values in Table II of Ref. \onlinecite{Nucara}. The  observation of the dip alone does not allow us \cite{Park} to give a meaningful determination of $\Gamma_D$ in sample B. } 
  
\begin{ruledtabular}
\begin{tabular}{cccccc}

Sample ($T$)  & $d$ (nm) & $\omega_p$ & $\Gamma_D$ & $\omega_{L3}$ & $\Gamma_{L3}$ \\

A (250 K) & 4 $\pm$ 1 & 980 $\pm$ 80 & 155 $\pm$ 20  & 784 & 35 	\\
A (10 K)  &  4 $\pm$ 1 & 950 $\pm$ 80   & 5 $\pm$ 2    & 784 & 23	\\
B (250 K) & 7 $\pm$ 2 & 3000 $\pm$ 500 &   & 787 & 27 	 \\
B (10 K)  &  7 $\pm$ 2 & 2900 $\pm$ 500 &      & 794 & 16	\\

\label{Table I}
\end{tabular}
\end{ruledtabular}
\end{table}


The quantities which describe the conducting properties of the q-2DES, as obtained from the fitting parameters in Table I, are reported in Table II. Therein, the electron surface density is  $n_s^{IR} = n^{IR}d = dm^* \epsilon_0 \omega_p^2/ e^2$ (where \cite{Dubroka} $m^* = 3.2m_e$ and $m_e$ is the bare electron mass). Both values turn out to be much lower under crystalline LAO (A) than under amorphous LAO (sample B). In sample A, $n_s$ is also lower than that measured in the carrier richest sample of Ref. \onlinecite{Dubroka}, but comparable with that reported for another sample in the same experiment. For sample A at both temperatures (10 and 250 K), the carrier mobility $\mu^{IR} = e/(2 \pi m^* c \Gamma)$ is in very good agreement with the dc mobility $\mu^{dc}$, also reported for comparison in Table II. It was extracted from the dc measurements of Ref. \onlinecite{Nucara} and from the present $n_s^{IR}$. 

\begin{figure} 
\begin{center}
\resizebox{1.0\columnwidth}{!}
{\includegraphics{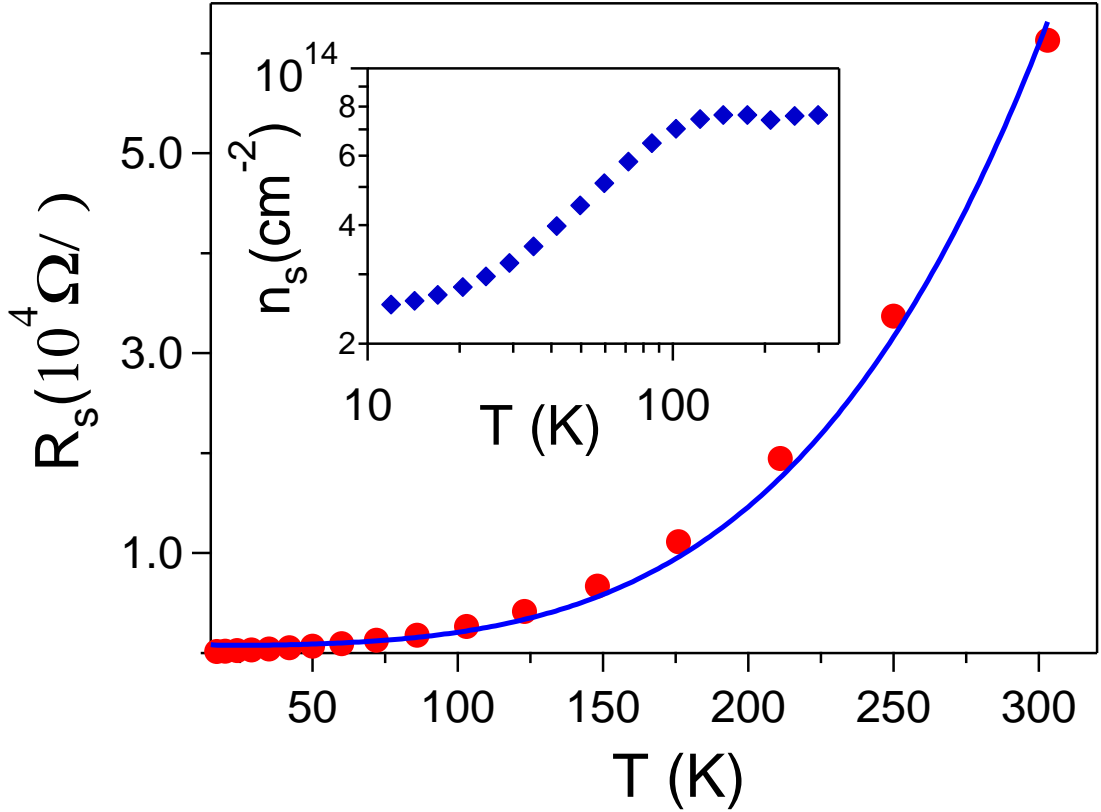}}
\caption{Sheet resistance $R$ (dots) of the q-2DES under amorphous LAO, measured on sample B, and fit (solid line) to the large-polaron model of Eq. \ref{R}. Inset: surface charge density $n_s$ vs. temperature as determined by Hall neasurements.} 
\label{Hall} 
\end{center}      
\end{figure}

In order to better understand the behavior of sample B, we have also performed on it dc transport and Hall measurements aimed at determining its electrodynamic properties vs. temperature at zero frequency. The results are shown in Fig. \ref{Hall}. In the inset, $n_s^{dc}$ is roughly constant between 300 and 100 K, to decrease by a factor of 3 between about 100 and 10 K. Such decrease is not observed in our mid-infrared spectra, and similar discrepancies were reported in Fig. 2 of Ref. \onlinecite{Dubroka} for LAO/STO, in Ref. \onlinecite{Dubroka17} again for LAO/STO, and in Ref. \onlinecite{Dubroka16} for $\gamma$-Al$_2$O$_3$/STO. Therein, the disagreement was explained  in terms of polaronic effects \cite{Calvani}. Indeed, the mobility values in Table II are consistent, \textit{e. g.},  with polaronic transport in TiO$_2$ \cite{Forro}. The weak localization observed in the q-2DES below 100 K in the inset of Fig. \ref{Hall} for amorphous LAO and in Fig. 2 of Ref. \onlinecite{Dubroka} for crystalline LAO points to polarons of the "large" type \cite{Devreese}. As usual, one can assume for the variable-temperature dependence of $n_s(T)$ an Arrhenius law \cite{Ziese} with a binding energy $\Delta$. Moreover, under the assumption of  scattering by acoustic phonons at not too high temperatures, the large-polaron mobility can be written \cite{Emin}

\begin{equation}
\mu(T) \propto eL^2(T_L/T)^4  
\label{mu}
\end{equation}

\noindent
where $L$ is the polaron dimension, and $T_L = \hbar c_s/(k_BL)$ with $c_s$ the velocity of sound in the solid.
The sheet resistance $R$ measured in sample B and reported in Fig. \ref{Hall} can then be fit by the Equation

\begin{equation}
R(T) \propto [n_s(T)e\mu(T)]^{-1} \propto T^4 exp(\Delta/2k_BT) 
\label{R}
\end{equation}

\noindent
The excellent fit to Eq. \ref{R} shown by a solid line in  Fig. \ref{Hall} provides for the polaronic charges a binding energy $\Delta \simeq$ 190 $k_B$. Therefore, with such a weak activation energy, the photoexcitation of the charges in the mid infrared will provide a $T$-independent $n_s^{IR}$, which should be compared with the $n_s^{dc}$ measured at saturation above 100 K (see the inset of Fig. \ref{Hall}). In Fig. \ref{Hall} this value is still lower than that in Table II by a factor of three, also possibly because it was taken after the Berreman experiments. Some sample degradation might be expected due to the thermal cycles needed for the infrared measurements and to the time elapsed. 

In Table II we have not reported $\mu^{IR}$ for sample B  because, following Ref. \onlinecite{Park}, the width of the Berreman dip only does not provide a meaningful estimate of $\Gamma_D$, the carrier relaxation rate. Moreover, when fitting the dip only, $n^{IR}$ and the q-2DES thickness $d$ are correlated, as shown  in Fig. 3 of the SI.  The large errors reported for these quantities in Table II have been obtained by building up at any $T$ a distribution of all their values which allowed us to obtain fits of equal accuracy to the $\Delta \Psi$ of sample B. The central values and standard deviations of those distributions have been reported in Table I (for $d$) and Table II (for $n^{IR}$ and, consequently, $n_s^{IR}$). Table II also displays the dc mobility extracted from  data in Fig. \ref{Hall} and in Fig. 1 of the SI \cite{SI}. One may notice that $\mu^{dc}$  is surprisingly high at low temperature, even better than in our crystalline sample A.


\begin{table} 
\caption{Surface density $n_s^{IR}$ and bulk density $n^{IR}$ of the q-2DES carriers at the interface between STO and crystalline (A) or amorphous (B) LAO at two temperatures, as extracted from the parameters of Table I with $m^* = 3.2 m_e$ \cite{Dubroka}. The carrier mobility $\mu^{IR}$ is extracted from the Berreman peak width $\Gamma_D$ (observed only in sample A). The dc mobility $\mu^{dc}$ is extracted from the data in Ref. \onlinecite{Nucara} for sample A, from the present Fig. \ref{Hall} and Fig. 1 of the SI \cite{SI} for sample B.} 
  
\begin{ruledtabular}
\begin{tabular}{cccccc}

Sample ($T$) & $n_s^{IR}$ (cm$^{-2}$)  & $n^{IR}$ (cm$^{-3}$) &  $\mu^{IR}$  (cm$^{2}$/Vs) &$\mu^{dc}$ (cm$^{2}$/Vs)  \\

A (250 K)  & (1.4 $\pm$ 0.3) $\times 10^{13}$ &    (3.4 $\pm$ 0.2) $\times 10^{19}$ & 20 $\pm$ 2 & 8 $\pm$ 1\\
A (10 K)    &  (1.3 $\pm$ 0.3) $\times 10^{13}$ &     (3.2 $\pm$ 0.2) $\times 10^{19}$ & 580 $\pm$ 40 & 560 $\pm$ 30 \\
B (250 K)  & (2.3 $\pm$ 0.7) $\times 10^{14}$ &   (3.2 $\pm$ 0.3) $\times 10^{20}$ &  & 10 $\pm$ 2	 	\\
B (10 K)    & (2.2 $\pm$ 0.7) $\times  10^{14}$  & (3.0 $\pm$ 0.3) $\times 10^{20}$ &  & 2200 $\pm$ 200	\\

\label{Table II}
\end{tabular}
\end{ruledtabular}
\end{table}

\begin{figure} 
\begin{center}
\resizebox{1.0\columnwidth}{!}
{\includegraphics{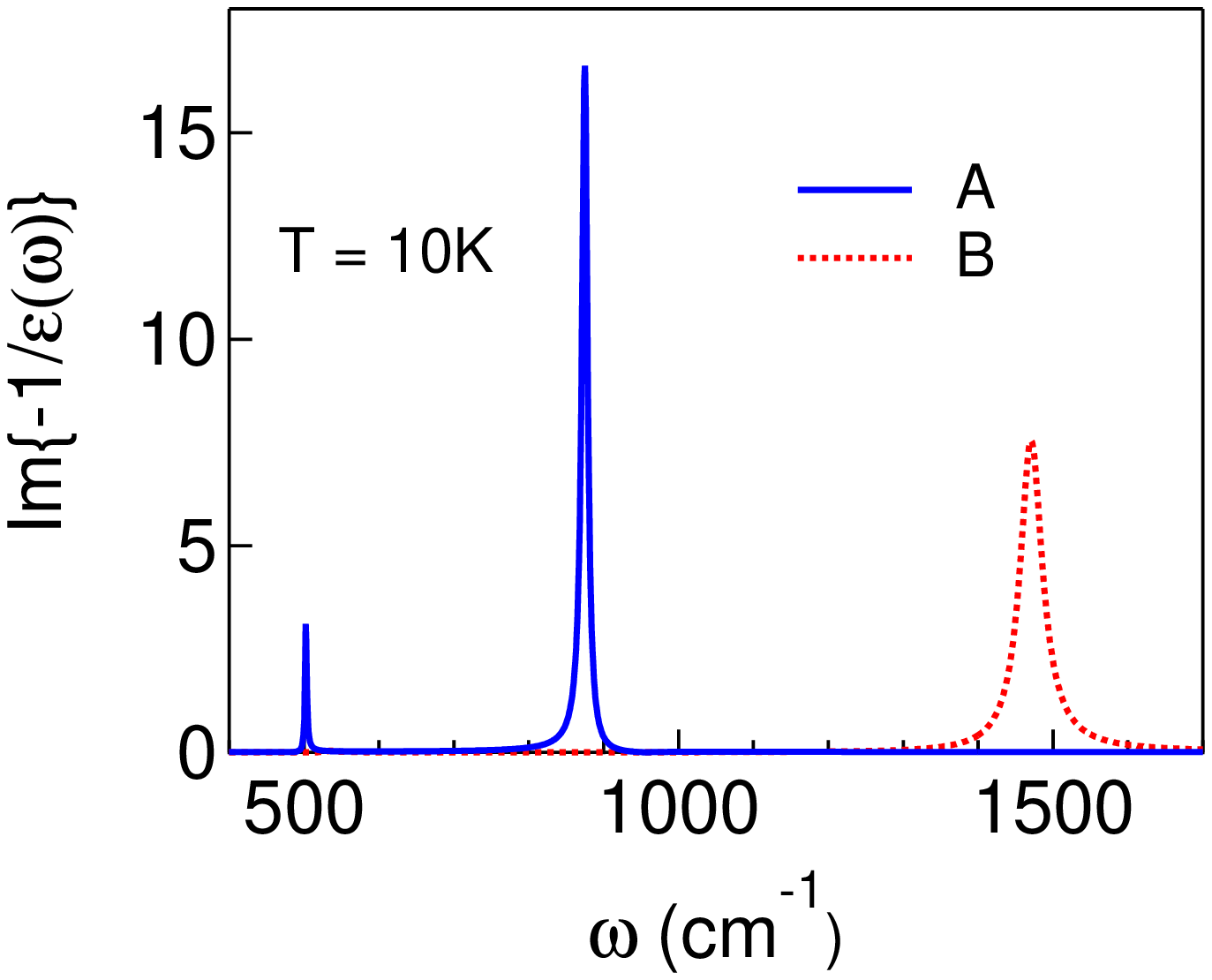}}
\caption{Energy-loss function calculated for both crystalline (A) and amorphous LAO (B) from the dielectric function which fits to data at 10 K. The peaks correspond to the Berreman resonances at $\omega_{Ber}$. Those at 502 (sample A) and 1479 cm$^{-1}$ (sample B) give rise to very weak features in the ellipsometric angle $\Psi$, not reported in the previous Figures.} 
\label{eloss} 
\end{center}      
\end{figure}

The lack of a Berreman peak close to the dip at 868 cm$^{-1}$ in sample B can be attributed to an insufficient signal-to-noise ratio where the reflectivity becomes too low, or to the charge density profile within the q-2DES\cite{Dubroka} ,which smears out the feature through a distribution of plasma frequencies. We can also tentatively explain that absence as follows. According to numerical simulations  of the BE reported in the literature \cite{Dubroka,Dubroka17}, while the dip frequency remains fixed,  the Berreman peak shifts rapidly to higher frequencies as $\omega_p$ increases. This is confirmed by Fig. \ref{eloss}, where the energy loss function $\Xi (\omega) = Im(-1/\tilde \epsilon(\omega)$ is plotted for both samples A and B as extrapolated from the $\tilde \epsilon_1$ of Eq.  \ref{LST+D}. We recall that the BE resonance, being a longitudinal spectral feature, is associated with a peak in $\Xi(\omega)$ \cite{Harbecke}. As one can see, while for sample A the peak corresponds exactly to that in $\Delta \Psi$ of Fig. \ref{crystal}, that of sample B should be displaced (in an ideal q-2DES system) to 1480 cm$^{-1}$, far away from the highest LO phonon of STO. Therein however,  $\Psi$  does not show any reproducible Berreman peak. One can tentatively assume that the peak is not observed because therein $R_p/R_s$ is low and the peak at $\omega{Ber}$ is much broadened by the disorder induced into the interface by amorphous LAO. One may also notice that a weaker peak in $\Xi(\omega)$ is predicted in the region of the LO2 phonon of sample A, shifted by 23  cm$^{-1}$ from $\omega_{L2}$. Neither this one has been observed. In general,  the weakness of the BE resonances in this q-2DES is not surprising if one recalls that, for a thin film, the optimum thickness to observe the BE is given by \cite{Harbecke}.

\begin{equation}
d_{opt} = (\lambda/2 \pi)(cos \theta/sin^2\theta)\Xi_{max}
\label{thick}
\end{equation}

\noindent
Replacing  in Eq. \ref{thick} our experimental values, one finds $d_{opt}$ = 37 nm for the q-2DES of crystalline LAO, and $d_{opt}$ = 70 nm for that of amorphous LAO. In both our samples (see Table II), the thickness $d$ of the q-2DES is smaller than $d_{opt}$ by an order of magnitude.

\section{Conclusion}

In conclusion, we have used grazing-angle infrared spectroscopy and the Berreman effect to investigate the electrodynamics of  the quasi-two-dimensional electron system (q-2DES) which forms spontaneously at the LaAlO$_3$/SrTiO$_{3}$  interface, both when the LAO film is crystalline (A) and when it is amorphous (B). In the latter case we have not observed the complete BE shape, and we have tentatively explained this result by a hardening and broadening of the Berreman peak caused by the unexpectedly high carrier density $n_s$ in sample B and by disorder at the interface, respectively. Nevertheless,  the BE has been shown once again to be a powerful tool to detect the electrodynamic response of ultrathin conducting films, in no-contact mode. By accurate fits to the BE resonances of both samples we have found that the thickness of the q-2DES is 4 $\pm$ 1 nm under crystalline LAO and 7 $\pm$ 2 nm under amorphous LAO .  Resistance and Hall measurements in sample B confirmed its high carrier density, even if the room temperature value of $n_s^{dc}$ is smaller than the previously measured $n_s^{IR}$ by about a factor of three. Moreover, unlike $n_s^{IR}$, $n_s^{dc}$ decreases below 100 K, as also reported in Ref. \onlinecite{Dubroka} for crystalline LAO, indicating that part of the charges originate from a  shallow localized state - possibly to be described as a large polaron - which, having an ionization energy of about 190 K, cannot affect the mid infrared spectra. Even if the q-2DES charge density in LAO/STO is known to strongly depend on the film growth conditions, it remains that it can assume quite high values also in the presence of amorphous LAO, where also the dc carrier mobility is surprisingly high.  These results may  be of interest for the present research effort aimed at better understanding the properties of the LAO/STO interface, and at exploiting them in the different applications that are being proposed for this intriguing system.

\acknowledgments
We wish to thank Adam Dubroka for illuminating discussions. This work has been partially supported by the Italian Ministry of University and Research through the PRIN project \textit{OXIDE} and by the European Commission CALIPSO Programme (HZBPHOTONS-CALIPSO-140) for the access to the synchrotron facilities.

\end{document}